\documentclass[12pt]{article} 
\usepackage{graphicx}
\usepackage{amsbsy}

\begin{document}
\baselineskip 10mm

\centerline{\large \bf Auxiliary-level-assisted operations
with charge qubits}
\centerline{\large \bf in semiconductors}

\vskip 6mm

\centerline{L. A. Openov}

\vskip 4mm

\centerline{\it Moscow Engineering Physics Institute (State
University), 115409 Moscow, Russia}

\vskip 6mm


\vskip 8mm

\centerline{\bf ABSTRACT}

We present a new scheme for rotations of a charge qubit associated with a
singly ionized pair of donor atoms in a semiconductor host. The logical
states of such a qubit proposed recently by Hollenberg {\it et al.} are
defined by the lowest two energy states of the remaining valence electron
localized around one or another donor. We show that an electron located
initially at one donor site can be transferred to another donor site
via an auxiliary molecular level formed upon the hybridization
of the excited states of two donors. The electron transfer is driven
by a single resonant microwave pulse in the case that the energies
of the lowest donor states coincide or two resonant pulses in the case that
they differ from each other. Depending on the pulse parameters, various
one-qubit operations, including the phase gate, the NOT gate, and the
Hadamard gate, can be realized in short times. Decoherence of an
electron due to the interaction with acoustic phonons is analyzed and shown
to be weak enough for coherent qubit manipulation being possible, at least
in the proof-of-principle experiments on one-qubit devices.

\newpage

\centerline{\bf 1. INTRODUCTION}

\vskip 2mm

Solid-state systems are of great interest in the search for a scalable
quantum computer technology. Several schemes for solid-state quantum
information processing have been proposed \cite{Barenco,Shnirman,Loss}. For
example, the coherent control of superconducting qubits \cite{Nakamura} and
their coupling \cite{Pashkin} have been demonstrated, the qubits being
encoded in the states of a Cooper-pair box. One promising area of current
investigation is concerned with the semiconductor-based devices.
In the Kane proposal \cite{Kane}, the qubits are defined by long-lived
nuclear spins of phosphorous dopants in a silicon host. They are manipulated
by external surface gates and RF magnetic fields. While long coherence times
of nuclear spins make the Kane scheme very promising, the single-spin
measurement remains a significant challenge \cite{Kane2}. This also concerns
an alternative Si:P architecture that uses electron spin states as qubits
\cite{Vrijen}.

Along with spin-based qubits, the charged-based qubits
in semiconductors are currently discussed as well.
The logical states of a charge semiconductor qubit may be formed by,
e. g., the ground and excited states of the electron in the single quantum
dot \cite{Barenco} or the spatially separated states of the electron
in two different quantum dots \cite{Openov,Valiev,Tanamoto,Oh,Hayashi}.
In spite of the
fact that decoherence of the charge-based qubits
is rather strong \cite{Fedichkin,Petta}, the charge
qubits are nevertheless believed to be realizable at the present
technological level due to their short operation times \cite{Hollenberg}.
One of the obstacles to the practical realization of scalable quantum
computation in the system of quantum dots is that it is extremely difficult,
if at all, to manufacture a set of quantum dots with identical
or at least predetermined characteristics each.
This complicates the issue, introducing the errors into the operations
with qubits \cite{Tsukanov} and generating a need for numerous ancillary
corrective gates. In this respect, it would be more reasonable to make use
of natural atoms (instead of "artificial" ones) as the localization
centers for the electrons carrying the quantum information. Recent advances
in manipulation with single atoms on the solid surface
\cite{Eigler} and atomically precise placement of single dopants in
semiconductors \cite{Schofield,Dzurak} make possible the construction of
rather complex solid-state atomic architectures.

Recently, Hollenberg {\it et al.} proposed a two-atom charge-qubit scheme
\cite{Hollenberg} and reported the first results on its fabrication and
characterization \cite{Dzurak} for the case of phosphorous dopants in
silicon. In that scheme, the buried donor charge qubit consists of two dopant
atoms $\sim 50$ nm apart in a semiconductor host. One of the donors is
singly ionized. The logical states are formed by the lowest two energy states
of the remaining valence electron localized at the left or the right donor,
$|0\rangle = |L\rangle$ and $|1\rangle = |R\rangle$, see Fig. 1. The qubit is
controlled by the surface
electrodes through adiabatic variations of the donor potentials.
Initialization and readout of the qubit are facilitated by a single
electron transistor. The coupling of such qubits via the Coulomb interaction
allows, in principle, to realize the conditional two-qubit gates
\cite{Hollenberg}.

It was shown in Ref. \cite{Hollenberg} that although the coherence time
$\tau_{coh}\sim 1$ ns for charge-based qubits is much shorter than for
their spin-based counterparts, the corresponding gate operations times are
also shorter, of order $\tau_{op}\sim 50$ ps. Note, however, that the ratio
of $\tau_{op}/\tau_{coh}\sim 10^{-1}$ seems to be insufficiently small
for the fault-tolerant scalable quantum computation
being possible \cite{DiVincenzo}. Here we propose an alternative scheme for
operations with buried donor charge qubits, instead of applying biases
to the surface gates. Our scheme is based on the effect of
electron transfer between the lowest states localized at different donors
upon the influence of a resonant pulse \cite{Openov}
or two resonant pulses \cite{Openov2}. Such a
transfer occurs via an excited molecular level of the double-donor system
and allows for implementation of different one-qubit rotations. The operation
times can be made orders of magnitude shorter than in the original proposal
\cite{Hollenberg}.

The paper is organized as follows. In Section II, we describe a three-level
model for the resonant electron transfer between the donors and briefly
discuss the relevant one-electron states of a P$_2^+$ molecular ion in Si.
Next we present the analytical solution for the unitary electron evolution
under the influence of microwave pulses. In Section III, we show that in the
P$_2^+$:Si system it is possible to realize various one-qubit operations,
including the NOT gate, the phase gate, and the Hadamard transformation.
Decoherence due to the electron interaction with acoustic phonons is studied
in Section IV. Discussion of the results is given in Section V.

\vskip 6mm

\centerline{\bf II. MODEL FOR THE RESONANT ELECTRON TRANSFER}

\vskip 2mm

We consider a singly ionized pair of phosphorous atoms embedded in silicon.
The remaining valence electron is described by the Hamiltonian
\begin{equation}
\hat{H}_0=\sum_{n} E_n |\chi_n\rangle \langle \chi_n| ~,
\label{H_0}
\end{equation}
where $E_n$ and $|\chi_n\rangle$ are, respectively, the one-electron
eigenenergies and eigenstates of the molecular ion P$_2^+$:Si. In general,
to calculate the energy spectrum and the wave functions
$\langle{\bf r}|\chi_n\rangle$ of the single-electron/double-donor system
beneath the surface, one should account for the conduction-band anisotropy,
the inter-valley terms, the surface effects, the potentials induced in the
substrate by the gate voltages, etc. This necessarily requires numerical
calculations, see, e. g. Ref. \cite{Kettle}. We note that although the
conduction-band edge of bulk Si has six degenerate minima, it has been shown
both experimentally \cite{Aggarwal} and theoretically \cite{Baldereschi} that
substitutional impurities break the translational symmetry of the crystal
lattice, thus lifting the degeneracy.
The spacing between the energy levels in the ground-state and
excited-states multiplets may be further increased through appropriate
choice of the gate potentials. Anyway, to quantify the structure of
the P$_2^+$:Si energy spectrum and wave functions, one should make
sophisticated numerical calculations for a specific donor configuration. In
this paper, however, we restrict ourselves to a semiquantitative
consideration based on an isotropic effective mass approximation
\cite{Barrett} that allows for an explicit analytical solution.
Then the problem reduces to that for the hydrogen-like molecular ion with
the effective Bohr radius $a_B^*\approx$ 3 nm and the effective Hartree unit
of energy $E^*=e^2/\varepsilon a_B^*\approx$ 40 meV, where $\varepsilon=11.7$
is the dielectric constant for silicon \cite{Note}. The energy spectrum of
the H$_2^+$ ion for different atomic separations is known with high accuracy
\cite{Bates}.

We approximate the Hamiltonian $\hat{H}_0$ in Eq. (\ref{H_0}) by the reduced
three-level Hamiltonian
\begin{equation}
\hat{H}_r=E_1|\chi_1\rangle\langle \chi_1|+E_2|\chi_2\rangle\langle \chi_2|+
E_{TR}|\chi_{TR}\rangle\langle \chi_{TR}| ~,
\label{H_r}
\end{equation}
where $|\chi_1\rangle$ and $|\chi_2\rangle$ are the lowest molecular states,
$1s\sigma_g$ and $2p\sigma_u$, whose wave functions are, respectively,
symmetric and antisymmetric about the midpoint of the line joining the
two donors, see Fig. 2, and $|\chi_{TR}\rangle$ is one of the excited
molecular states discussed below. It is convenient to go from the states
$|\chi_1\rangle$ and $|\chi_2\rangle$ delocalized over the P$_2^+$:Si ion to
the states $|L\rangle=\bigl[|\chi_1\rangle+|\chi_2\rangle\bigr]/\sqrt{2}$ and
$|R\rangle=\bigl[|\chi_1\rangle-|\chi_2\rangle\bigr]/\sqrt{2}$ localized at
the left and the right donor, respectively. For donor separations $R>>a_B^*$,
the wave functions $\langle{\bf r}|L\rangle$ and $\langle{\bf r}|R\rangle$
are almost indistinguishable from the one-electron $1s$-orbitals of the
corresponding donor atoms.

The states $|L\rangle$ and $|R\rangle$ form the qubit logical states
$|0\rangle$ and $|1\rangle$, respectively. These states are well defined if
the thermal energy $k_BT$ is much lower than the differences
$\Delta E_{31}=E_3-E_1$ and $\Delta E_{32}=E_3-E_2$ between the energy
$E_3$ of the excited molecular state $|\chi_3\rangle$ and the energies $E_1$
and $E_2$, respectively. At $R>>a_B^*$ one has $E_1\approx E_2\approx -E^*/2$
and $E_3\approx -E^*/8$, so that
$\Delta E_{31}\approx\Delta E_{32}\approx 3E^*/8\approx 15$ meV. Since the
states $|L\rangle$ and $|R\rangle$ are not the exact eigenstates of the
Hamiltonian $\hat{H}_r$, in the absence of external fields the initial qubit
state $|\Psi(0)\rangle=\alpha|L\rangle + \beta|R\rangle$ will evolve with
time as
\begin{equation}
|\Psi (t)\rangle = e^{-i\hat{H}_r t/\hbar} |\Psi (0)\rangle =
e^{-iE_1 t/\hbar}\biggl\{ |\Psi (0)\rangle +
i(\beta-\alpha)e^{-i\Delta E_{21}t/2\hbar}
\sin(\Delta E_{21}t/2\hbar)\biggl[|L\rangle-|R\rangle\biggr]\biggr\} ~,
\label{Psi}
\end{equation}
where $\Delta E_{21}=E_2-E_1$. Note that at $t<<t_0=\hbar/\Delta E_{21}$
the initial qubit state remains almost unchanged (not counting the common
phase). Since at $x=R/a_B^*>>1$, the value of $\Delta E_{21}$ is
exponentially small \cite{Bardsley,Scott},
\begin{equation}
\frac{\Delta E_{21}}{E^*}=4x e^{-x-1}\left [
1+\frac{1}{2x}+O\left(\frac{1}{x^2}\right)\right ] ~,
\label{DeltaE_21}
\end{equation}
the period
$t_0\sim\hbar/\Delta E_{21}$ it takes for the qubit state to change is rather
long, $t_0 > 1$ $\mu$s at $R>60$ nm. In what follows we shall consider the
processes taking place in time intervals much shorter than $t_0$ and hence
ignore the off-diagonal term
$\bigl[-(\Delta E_{21}/2)|L\rangle \langle R| + h.c. \bigr]$ in
$\hat{H}_r$ that gives rise to the electron tunneling
$|L\rangle\rightleftharpoons |R\rangle$. Then the Hamiltonian (\ref{H_r})
takes the form
\begin{equation}
\hat{H}_r\approx\frac{E_1+E_2}{2}
\biggl[|L\rangle\langle L|+|R\rangle\langle R|\biggr]
+ E_{TR}|\chi_{TR}\rangle\langle \chi_{TR}| ~,
\label{H_r2}
\end{equation}
where $(E_1+E_2)/2\approx E_1\approx E_2$ at $R>>a_B^*$.
In the general case that the qubit is
biased by the gate voltages, the energies $E_L$ and $E_R$ of the lowest
states localized, respectively, at the left and the right donor differ
from each other. In this case, the localized states are all the more good
approximations to the energy eigenstates, and the Hamiltonian $\hat{H}_r$
reads
\begin{equation}
\hat{H}_r\approx E_L|L\rangle\langle L|+E_R|R\rangle\langle R|+
E_{TR}|\chi_{TR}\rangle\langle \chi_{TR}| ~.
\label{H_r3}
\end{equation}

Now let the buried donor charge qubit interact with an external
electromagnetic field ${\bf E}(t)$. Then the Hamiltonian becomes
\begin{equation}
\hat{H}(t)=\hat{H}_r+\hat{V}(t)~,
\label{H}
\end{equation}
where the interaction term $\hat{V}(t)$ is
\begin{equation}
\hat{V}(t)={\bf E}(t)\biggl[ {\bf d}_L|\chi_{TR}\rangle \langle L|
+ {\bf d}_R|\chi_{TR}\rangle \langle R| + h.c. \biggr] ~,
\label{V}
\end{equation}
with ${\bf d}_L=\langle \chi_{TR}|-e{\bf r}|L\rangle$ and
${\bf d}_R=\langle \chi_{TR}|-e{\bf r}|R\rangle$ being the electric dipole
moments for the transitions $|L,R\rangle \rightleftharpoons
|\chi_{TR}\rangle$ between, respectively, the localized states $|L\rangle$
and $|R\rangle$ and one of the excited molecular states $|\chi_{TR}\rangle$
delocalized over the double-donor system. For definiteness, we choose this
state to be the third one-electron state $|\chi_3\rangle$ of the molecular
ion P$_2^+$:Si. At $E_L=E_R$ and
$R/a_B^*> 6$, this is the $3d\sigma_g$ state whose wave
function $\langle {\bf r}|\chi_3\rangle$ is symmetric about the midpoint of
the line joining the two donors and has its maxima at the donor
locations \cite{Fudzinaga}, see Fig. 2. If the donors are arranged along
the $x$-axis, the state
$|\chi_3\rangle$ is formed upon the hybridization of $|2S\rangle_{L,R}$ and
$|2P_x\rangle_{L,R}$ atomic states of the donors, and
$\langle {\bf r}|\chi_3\rangle$ in the vicinity of the left/right donor
equals to $\bigl[\langle {\bf r}|2S\rangle_{L,R}
\mp \langle {\bf r}|2P_x\rangle_{L,R}\bigr]/2$ at $R>>a_B^*$. Note that for
such a choice of the state $|\chi_3\rangle$, the electric field should have a
nonzero $x$-component in order that ${\bf d}_{L,R}\neq 0$.

We consider two cases:
(a) $E_L=E_R\approx E_1$ and (b) $E_L\neq E_R$, the desired value of the
difference $E_R-E_L$ being discussed below. In the case (a), we suppose
${\bf E}(t)$ to oscillate at a frequency
$\omega = (E_{TR}-E_{L,R})/\hbar$,
\begin{equation}
{\bf E}(t)={\bf E_0}(t)\cos(\omega t)~,
\label{Ea}
\end{equation}
where ${\bf E_0}(t)$ is the slowly varying envelope of the field. Making use
of the resonant approximation \cite{NOTE2}, i. e., omitting the rapidly
oscillating terms with the frequencies
$\pm (\omega +E_{TR}/\hbar-E_{L,R}/\hbar)$ from
the Hamiltonian, we have
\begin{equation}
\hat{V}(t)=\frac{1}{2}e^{-i\omega t}\biggl[ \lambda_L(t)
|\chi_{TR}\rangle \langle L| + \lambda_R(t)|\chi_{TR}\rangle
\langle R| \biggr] + h. c. ~,
\label{Va}
\end{equation}
where $\lambda_{L,R}(t)={\bf E_0}(t){\bf d}_{L,R}$. In the case (b),
the field ${\bf E}(t)$ has two components oscillating at frequencies
$\omega_L=(E_{TR}-E_L)/\hbar$ and $\omega_R=(E_{TR}-E_R)/\hbar$,
\begin{equation}
{\bf E}(t)={\bf E_{01}}(t)\cos(\omega_L t) +
{\bf E_{02}}(t)\cos(\omega_R t+\phi)~,
\label{Eb}
\end{equation}
where $\phi$ is the phase shift between the two components. In the resonant
approximation \cite{NOTE2} one has
\begin{equation}
\hat{V}(t)=\frac{1}{2}e^{-i\omega_L t}\lambda_L(t)
|\chi_{TR}\rangle \langle L| + \frac{1}{2}e^{-i\omega_R
t-i\phi}\lambda_R(t) |\chi_{TR}\rangle \langle R| + h. c.~,
\label{Vb}
\end{equation}
where $\lambda_{L,R}(t)={\bf E_{01,2}}(t){\bf d}_{L,R}$.
In this paper, we restrict ourselves to the rectangular pulse shape, so
that ${\bf E_0}(t)$ in Eq. (\ref{Ea}) and both ${\bf E_{01}}(t)$ and
${\bf E_{02}}(t)$ in Eq. (\ref{Eb}) are constant at $0<t<\tau_{op}$ and zero
elsewhere.

It is straightforward to solve the non-stationary Schr\"{o}dinger equation
for the state vector $|\Psi (t)\rangle$,
\begin{equation}
i\hbar\frac{\partial |\Psi(t)\rangle}{\partial t}=\hat{H}(t)|\Psi(t)\rangle~,
\label{Sch(t)}
\end{equation}
with the Hamiltonian $\hat{H}(t)$ in Eq. (\ref{H}) given by Eqs. (\ref{H_r2})
and (\ref{Va}) in the case (a) or (\ref{H_r3}) and (\ref{Vb})
in the case (b), and to find the coefficients
$C_L(t)$, $C_R(t)$, and $C_{TR}(t)$ in the expansion of $|\Psi(t)\rangle$ in
terms of the states $|L\rangle$, $|R\rangle$, and $|\chi_{TR}\rangle$,
\begin{equation}
|\Psi(t)\rangle = C_L(t)e^{-iE_L t/\hbar}|L\rangle +
C_R(t)e^{-iE_R t/\hbar}|R\rangle +
C_{TR}(t)e^{-iE_{TR} t/\hbar}|\chi_{TR}\rangle ~,
\label{Psi(t)}
\end{equation}
provided that $|\Psi(0)\rangle = \alpha|L\rangle + \beta|R\rangle$, where
$|\alpha|^2+|\beta|^2=1$. In the case (a) we have
\begin{eqnarray}
&&C_L(t)=\alpha\left [
1-\frac{2|\lambda_L|^2}{|\lambda_L|^2+|\lambda_R|^2} \sin^2(\Omega
t)\right ] -\beta\frac{2\lambda_L^*\lambda_R}
{|\lambda_L|^2+|\lambda_R|^2}\sin^2(\Omega t)~, \nonumber \\
&&C_R(t)=-\alpha\frac{2\lambda_L\lambda_R^*}{|\lambda_L|^2+|\lambda_R|^2}
\sin^2(\Omega t)+\beta\left [ 1-\frac{2|\lambda_R|^2}
{|\lambda_L|^2+|\lambda_R|^2}\sin^2(\Omega t) \right ] ~,
\nonumber
\\ &&C_{TR}(t)=-i\frac{\alpha\lambda_L+\beta\lambda_R}
{\sqrt{|\lambda_L|^2+|\lambda_R|^2}}\sin(2\Omega t)~,
\label{C(t)}
\end{eqnarray}
where
\begin{equation}
\Omega=\frac{\sqrt{|\lambda_L|^2+|\lambda_R|^2}}{4\hbar}~.
\label{Omega}
\end{equation}
In the case (b), the coefficients $C_L(t)$, $C_R(t)$, and $C_{TR}(t)$ are
also given by Eqs. (\ref{C(t)}) and (\ref{Omega}) with the only exception
that $\lambda_R$ should be replaced by $\lambda_R e^{-i\phi}$.
From Eqs. (\ref{C(t)}) and (\ref{Omega}) one can see that at
$t=\tau_{op}=\pi k/2\Omega$ (hereafter $k$ is a positive integer) the
coefficient $C_{TR}$ vanishes, so that the state vector $|\Psi(t)\rangle$
remains in the qubit subspace $\{|L\rangle,|R\rangle\}$ and
$|C_L(\tau_{op})|^2+|C_R(\tau_{op})|^2=1$.
In particular, if $C_L(0)=1$ and $C_R(0)=0$, then $C_L(\tau_{op})=0$ and
$C_R(\tau_{op})=\pm 1$ at $\lambda_L=\mp\lambda_R$ and odd $k$,
i. e., there is a complete population transfer
$|L\rangle\rightarrow|R\rangle$, see Ref. \cite{Openov}.
So, the auxiliary excited state
$|\chi_{TR}\rangle$ plays the role of the "transport" state, in that it
assists the qubit evolution by means of the electron transfer between the
states $|L\rangle$ and $|R\rangle$ as the pulse is on but remains unpopulated
after the pulse is off.

\vskip 6mm

\centerline{\bf III. QUBIT ROTATIONS}

\vskip 2mm

In this Section we show that the auxiliary-state-assisted electron transfer
between the two donors allows for various qubit rotations. In the case (a)
that the two donors in the molecular ion P$_2^+$:Si are equivalent, i. e.,
$E_L=E_R$ and $|\lambda_L|=|\lambda_R|$, the qubit state $|\Psi(t)\rangle$
at the operation time $\tau_{op}$ remains unchanged,
\begin{equation}
|\Psi(\tau_{op})\rangle = e^{-iE_L\tau_{op}/\hbar}|\Psi(0)\rangle~,
\label{Psi(tau_op)a1}
\end{equation}
if $\tau_{op}=\pi k/\Omega$, or changes into
\begin{equation}
|\Psi(\tau_{op})\rangle = \pm e^{-iE_L\tau_{op}/\hbar}
\biggl[\beta|L\rangle + \alpha|R\rangle \biggr] ~,
\label{Psi(tau_op)a2}
\end{equation}
if $\tau_{op}=\pi (2k-1)/2\Omega$ and $\lambda_L=\mp\lambda_R$, see
Eqs. (\ref{Psi(t)}) and (\ref{C(t)}). The latter
corresponds to the quantum NOT operation.

The case (b) seems to be more realistic because of the different local
atomic surroundings of the donors in the pair due to both the uncontrollable
damage of the host upon ion implantation and the probabilistic variations in
the path taken through the substrate by each implanted ion \cite{Dzurak}.
Besides, the surface gates can be used to intentionally tune $E_L$ and $E_R$
to the predetermined values. Moreover, one can change the values of
$\lambda_L$ and $\lambda_R$ separately through the changes in the
electric field amplitudes $E_{01}$ and $E_{02}$. It follows from
Eqs. (\ref{Psi(t)}) and (\ref{C(t)}) that the relative phase shift operation
is implemented at $\tau_{op}=\pi k/\Omega$,
\begin{equation}
|\Psi(\tau_{op})\rangle = e^{-iE_L\tau_{op}/\hbar}
\biggl[\alpha|L\rangle +
\beta e^{-i(E_R-E_L)\tau_{op}/\hbar}|R\rangle \biggr] ~,
\label{Psi(tau_op)b1}
\end{equation}
while the value of $\tau_{op}=\pi (2k-1)/2\Omega$ corresponds to realization
of the quantum NOT operation,
\begin{equation}
|\Psi(\tau_{op})\rangle = \pm e^{-iE_L\tau_{op}/\hbar-i\phi}
\biggl[\beta|L\rangle + \alpha|R\rangle \biggr]~,
\label{Psi(tau_op)b2}
\end{equation}
if $\lambda_L=\mp\lambda_R$ and $\phi=\pi n + (E_R-E_L)\tau_{op}/2\hbar$
(hereafter $n$ is an integer), or to the Hadamard transformation,
\begin{equation}
|\Psi(\tau_{op})\rangle = \pm e^{-iE_L\tau_{op}/\hbar}
\left[\frac{\alpha+\beta}{\sqrt{2}}|L\rangle +
\frac{\alpha-\beta}{\sqrt{2}}|R\rangle\right]~,
\label{Psi(tau_op)b3}
\end{equation}
if $(E_R-E_L)\tau_{op}/\hbar=2\pi m$ (where $m$ is a positive integer). Here
the plus sign corresponds to the values of $\phi=2\pi n$ and
$\lambda_L=-\lambda_R(\sqrt{2}-1)$ or $\phi=\pi(2n+1)$ and
$\lambda_L=\lambda_R(\sqrt{2}-1)$, and the minus sign corresponds to the
values of $\phi=2\pi n$ and $\lambda_L=\lambda_R(\sqrt{2}+1)$ or
$\phi=\pi(2n+1)$ and $\lambda_L=-\lambda_R(\sqrt{2}+1)$.

So, various one-qubit operations can be implemented on the buried donor
charge qubit through appropriate choices of the pulse frequency, phase,
amplitude, and duration. Let us estimate the value of the operation time
$\tau_{op}\sim 1/\Omega\sim \hbar/|\lambda_{L,R}|\sim \hbar/ea_B^*E_0$, see
Section II. For the field amplitude $E_0\sim 1$ V/cm one has
$\tau_{op}\sim 1$ ns. Increase in the pulse intensity will cause the value of
$\tau_{op}$ to decrease down to the picosecond time scale, so that the
value of $\tau_{op}$ can be made orders of magnitude shorter than the period
$t_0$ it takes for the qubit state to change due to the direct electron
tunneling $|L\rangle\rightleftharpoons |R\rangle$, see Section II, as well as
the operation times in the case that the qubit is manipulated by
adiabatically varying the potentials of the surface gates \cite{Hollenberg}.
Note that in the case (b) the energies $E_L$ and $E_R$ should be sufficiently
different from
each other in order all these operations could be implemented in short times
to avoid decoherence, as discussed below. For example,
at $\tau_{op}\sim 1$ ps one should have $E_R-E_L\sim 3$ meV.

\vskip 6mm

\centerline{\bf IV. DECOHERENCE EFFECTS}

\vskip 2mm

The uncontrolled interaction of the quantum system with its environment leads
to entanglement between the states of the system and the environmental
degrees of freedom. This disturbs the unitary evolution of the system and
results in the loss of coherence. There are various sources of decoherence
in solids. For the charge qubit considered in this paper, the decoherence
due to the phonon emission/absorption processes was studied in
Refs. \cite{Hollenberg,Barrett} and found to be much weaker than the
decoherence due to both Nyquist-Johnson voltage fluctuations in the surface
electrodes and 1/f noise from the background charge fluctuations. Note,
however, that there are two mechanisms of the phonon-induced decoherence
which are caused by either the energy relaxation processes or the
virtual-phonon dephasing processes. Which one of those mechanisms is
dominant, depends on the specific parameters of the quantum system and its
environment, as well as on the operation times. Here we show that the
dephasing processes play a decisive role in limiting the fault tolerance
of the buried donor charge qubit. For
simplicity, we consider the qubit at zero temperature and assume isotropic
acoustic phonons with the linear dispersion law, $\omega_{{\bf q}}=sq$,
where $s$ is the speed of sound.

First we recall some general concepts concerning the transition probability
for an electron moving in the time-dependent potential. If an electron, being
initially in the state $|i\rangle$ of the discrete energy spectrum, interacts
with the harmonic field
\begin{equation}
\hat{V}(t)=\hat{F} e^{-i\omega t}+\hat{F}^+ e^{i\omega t}~,
\label{Vharm}
\end{equation}
then the probability amplitude to find it in the state $|f\rangle$ at a time
$t$ is given by the following expression that results from the first-order
perturbation theory \cite{Landau},
\begin{equation}
a_{i\rightarrow f}(\omega,t)=
F_{fi}\frac{e^{-i(\omega_{if}+\omega)t}-1}{\hbar(\omega_{if}+\omega)}+
F_{if}^*\frac{e^{-i(\omega_{if}-\omega)t}-1}{\hbar(\omega_{if}-\omega)}~,
\label{aif}
\end{equation}
where $\omega_{if}=(E_i-E_f)/\hbar$. The common approach is to
ignore the first term in Eq. (\ref{aif}) and make use of the
expression
\begin{equation}
\lim_{t\rightarrow\infty}\frac{\sin^2(\epsilon t)}{\pi t\epsilon^2}=
\delta(\epsilon)~,
\label{delta}
\end{equation}
thus arriving at the so called Fermi golden rule for the transition
probability,
\begin{equation}
W_{i\rightarrow f}(\omega,t)=|a_{i\rightarrow f}(\omega,t)|^2\approx
|F_{if}|^2\frac{4\sin^2\left(\frac{\omega_{if}-\omega}{2}t\right)}
{\hbar^2(\omega_{if}-\omega)^2}\approx
\frac{2\pi}{\hbar}|F_{if}|^2\delta(\hbar\omega_{if}-\hbar\omega)t\equiv
\Gamma_{i\rightarrow f}(\omega)t~,
\label{Wif(omega,t)}
\end{equation}
where $\Gamma_{i\rightarrow f}(\omega)$ is the time-independent transition
rate. The $\delta$-function
reflects the energy conservation, $\hbar\omega_{if}=\hbar\omega$, for such a
transition.

Electron-phonon coupling in confined systems is described by the Hamiltonian
\begin{equation}
\hat{H}_{el-ph}=\sum_{\bf q}\lambda({\bf q})\hat{\rho}({\bf q})
\left[\hat{b}_{\bf q}^+ +\hat{b}_{-{\bf q}}^{}\right]~,
\label{Helph}
\end{equation}
where $\hat{b}_{\bf q}^+$ and $\hat{b}_{\bf q}^{}$ are, respectively, the
operators of creation and annihilation of a phonon with the wave vector
${\bf q}$,
$\hat{\rho}({\bf q})=\int{d{\bf r} e^{i{\bf qr}}\hat{\rho}({\bf r})}$
is the Fourier transform of the electron density operator
$\hat{\rho}({\bf r})=\sum_{mn}\Psi_m^*({\bf r})\Psi_n^{}({\bf r})
|m\rangle\langle n|$, and $\lambda({\bf q})$ is the microscopic
electron-phonon interaction matrix element that can be expressed in terms of
the deformation potential $D$ and the density of the crystal $\rho$ as
\begin{equation}
\lambda({\bf q})=
qD\left(\frac{\hbar}{2\rho\omega_{\bf q}\Omega}\right)^{1/2}~,
\label{lambda}
\end{equation}
with $\Omega$ being the normalizing volume. If the harmonic
field (\ref{Vharm}) is associated with a
deformation phonon having the frequency $\omega_{{\bf q}}$, then, taking into
account that the deformation fields produced by the phonons with different
wave vectors are not correlated,  one has for the total transition rate
\cite{Bockelmann,NOTE3}
\begin{equation}
\Gamma_{i\rightarrow f}=\frac{2\pi}{\hbar}\sum_
{{\bf q}}|F_{if}({\bf q})|^2\delta(\hbar\omega_{if}-\hbar\omega_{\bf q})~,
\label{Gammaif}
\end{equation}
where
\begin{equation}
F_{if}({\bf q})=\lambda({\bf q})\langle i|e^{i{\bf qr}}|f\rangle~.
\label{Fif}
\end{equation}

\newpage

\centerline{\bf A. Decoherence during adiabatic variations of the surface
gate potentials}

\vskip 2mm

In the case that the buried donor charge qubit is controlled by the surface
gates \cite{Hollenberg}, so that the state vector $|\Psi(t)\rangle$
remains in the qubit
subspace $\{|L\rangle , |R\rangle\}$ during the operation, and
the overlap $\langle L|R\rangle$ is negligibly small, the Hamiltonian
(\ref{Helph}) can be written in the spin-boson form \cite{Brandes},
\begin{equation}
\hat{H}_{el-ph}=\hat{\sigma}_z\sum_{\bf q}g({\bf q})
\left[\hat{b}_{\bf q}^+ +\hat{b}_{-{\bf q}}^{}\right]~,
\label{Helph2}
\end{equation}
where $\hat{\sigma}_z = |L\rangle\langle L|-|R\rangle\langle R|$ and
\begin{equation}
g({\bf q})=\frac{\lambda({\bf q})}{2}
\left[\langle L|e^{i{\bf qr}}|L\rangle - \langle R|e^{i{\bf qr}}|R\rangle
\right]~.
\label{g(q)}
\end{equation}
Since $\langle{\bf r}|L,R\rangle=(\pi(a_B^*)^3)^{-1/2}
\exp(-|{\bf r}-{\bf r}_{L,R}|/a_B^*)$ for $1s$-orbitals, where
${\bf r}_{L,R}$ are the donor coordinates, one has \cite{Fedichkin}
\begin{equation}
g({\bf q})=
-i\lambda({\bf q})\frac{\sin(q_xR/2)}{\bigl[1+(qa_B^*)^2/4\bigr]^2}~,
\label{g(q)2}
\end{equation}
where $q_x$ is the component of the phonon wave vector along the line joining
the two donors, and we chose the origin of the coordinates in between the
donors, so that ${\bf r}_{L,R}=\mp(R/2){\bf e}_x$.

Fedichkin and Fedorov \cite{Fedichkin} have shown that at $T=0$ decoherence
upon implementing the phase operation emerges as pure dephasing, the electron
density matrix being given by the general expression \cite{Kampen,Mozyrsky},
\begin{eqnarray}
\left(
\begin{array}{cccc}
\rho_{LL}(0) &\ \rho_{LR}(0) e^{-B^2(t)+i(E_R-E_L)t/\hbar} \\
\rho_{RL}(0) e^{-B^2(t)-i(E_R-E_L)t/\hbar} &\ \rho_{RR}(0) \\
\end{array}
\right)~,
\label{rho}
\end{eqnarray}
with the spectral function
\begin{equation}
B^2(t)=\frac{8}{\hbar^2}\sum_{{\bf q}}
\frac{|g({\bf q})|^2}{\omega_{{\bf q}}^2}
\sin^2\left(\frac{\omega_{{\bf q}}t}{2}\right)~.
\label{B2(t)}
\end{equation}
There is no relaxation in this case since in order the phase operation could
be implemented, the energies $E_L$ and $E_R$ should be sufficiently different
from each other \cite{Fedichkin}, so that the basis
$\{|L\rangle , |R\rangle\}$ coincides with the energy basis of the electron
in the double donor system, and the electron term (\ref{H_r3}) commutes with
the interaction term (\ref{Helph2}) in the Hamiltonian. As a result, the
diagonal elements of the density matrix remain unchanged. On the other hand,
decoherence upon implementing the quantum NOT operation (where $E_L=E_R$ and
the energy basis of the electron is formed by the states
$|\chi_{1,2}\rangle=\bigl[|L\rangle\pm|R\rangle\bigr]/\sqrt{2}$,
see Section II) was suggested to be caused by relaxation \cite{Fedichkin},
so that both off-diagonal and diagonal elements of the density matrix
decrease exponentially with time, the relaxation rate
$\Gamma_{2\rightarrow 1}$ being \cite{Fedichkin,Barrett}, see
Eq. (\ref{Gammaif}),
\begin{equation}
\Gamma_{2\rightarrow 1}=\frac{D^2}{4\pi\rho\hbar s^2}
\frac{q_{21}^3}{{\bigl[1+(q_{21}a_B^*)^2/4\bigr]^4}}
\left(1-\frac{\sin(q_{21}R)}{q_{21}R}\right)~,
\label{Gamma21}
\end{equation}
where $q_{21}=\Delta E_{21}/s\hbar$, see Eq. (\ref{DeltaE_21}).

Note, however, that the approximation (\ref{Wif(omega,t)}) for
$W_{i\rightarrow f}(\omega,t)$ and, accordingly, the equation (\ref{Gammaif})
for $\Gamma_{i\rightarrow f}$ are valid provided the time $t$ is sufficiently
long, see Eq. (\ref{delta}). To quantify the applicability of this
approximation, let us analyze the more general expression for
$W_{i\rightarrow f}(t)$ that follows from Eq. (\ref{Wif(omega,t)}),
\begin{equation}
W_{i\rightarrow f}(t)=\frac{4}{\hbar^2}\sum_{{\bf q}}
|F_{if}({\bf q})|^2
\frac{\sin^2\left(\frac{\omega_{if}-\omega_{\bf q}}{2}t\right)}
{(\omega_{if}-\omega_{\bf q})^2} ~.
\label{Wif(t)}
\end{equation}
One can roughly distinguish two phonon contributions to
$W_{i\rightarrow f}(t)$, one being from the "resonant component", i. e.,
from the $\delta$-function-like peak of
$\sin^2(\frac{\omega_{if}-\omega_{\bf q}}{2}t)/
(\omega_{if}-\omega_{\bf q})^2$ as a function of $q$ at
$q=q_{if}=\omega_{if}/s$, with the height $t^2/4$ and the width $\sim 1/st$,
and another from the remaining "non-resonant background" of the phonon
spectrum. The former can be estimated as
\begin{equation}
W_{i\rightarrow f}^{(1)}(t)\sim\frac{\Omega q_{if}^2}{\hbar^2 s}
|F_{if}(q_{if})|^2 t ~,
\label{Wif1(t)}
\end{equation}
and the latter as
\begin{equation}
W_{i\rightarrow f}^{(2)}(t)\sim\frac{\Omega \Delta q}{\hbar^2 s^2}
|F_{if}(q_{max})|^2
\label{Wif2(t)}
\end{equation}
at $q_{if}<<q_{max}$ and
\begin{equation}
W_{i\rightarrow f}^{(2)}(t)\sim\frac{\Omega \Delta q}
{\hbar^2 s^2}\left(\frac{q_{max}}{q_{if}}\right)^2|F_{if}(q_{max})|^2
\label{Wif2'(t)}
\end{equation}
at $q_{if}>>q_{max}$,
where $q_{max}$ is the wave vector at which the function $|F_{if}(q)|^2$ has
a maximum, and $\Delta q$ is a characteristic width of $|F_{if}(q)|^2$ in the
maximum. The specific values of $\Delta q$, $q_{max}$, and $F_{if}(q_{max})$
depend on the specific type of wave functions $\langle{\bf r}|i\rangle$ and
$\langle{\bf r}|f\rangle$ in the matrix element
$\langle i|e^{i{\bf qr}}|f\rangle$. Now if, e. g., $q_{if}<<q_{max}$ and
we are interested in the
transition probability $W_{i\rightarrow f}(t)$ at a moment of time $t$ such
that $sq_{if}^2|F_{if}(q_{if})|^2t<<\Delta q|F_{if}(q_{max})|^2$, then
$W_{i\rightarrow f}^{(1)}(t)<<W_{i\rightarrow f}^{(2)}(t)$, and hence the
Fermi golden rule appears to be broken \cite{Liu,Openov3}.
This is due to violation
of the energy conservation at short times \cite{Landau}.

An inspection of the phonon-induced transitions between the states
$|\chi_{1,2}\rangle=\bigl[|L\rangle\pm|R\rangle\bigr]/\sqrt{2}$ of the double
donor system with $E_L=E_R$ and the donor separation $R>>a_B^*$ (these
transitions are relevant for decoherence during the implementation of the NOT
operation \cite{Fedichkin}) provides an
illustrative example of the departure from the Fermi golden rule. In this
case $\langle 2|e^{i{\bf qr}}|1\rangle=
\bigl[\langle L|e^{i{\bf qr}}|L\rangle-
\langle R|e^{i{\bf qr}}|R\rangle\bigr]/2$, so that
$F_{21}({\bf q})=g({\bf q})$, see Eq. (\ref{g(q)2}), and the resonant
component of the transition probability is
$W_{2\rightarrow 1}^{(1)}(t)\sim q_{21}^3D^2t/\rho\hbar s^2$, in accordance
with the value of the relaxation rate $\Gamma_{2\rightarrow 1}$ given by
Eq. (\ref{Gamma21}). Since the
value of $q_{21}=\Delta E_{21}/\hbar s$ decreases exponentially with $R$,
see Eq. (\ref{DeltaE_21}), the value of $\Gamma_{2\rightarrow 1}$ decreases
exponentially as well, going below 10$^3$ s$^{-1}$ at $R/a_B^*>10$,
see Fig. 5 in Ref. \cite{Barrett}. On the other hand, since
$q_{21}<<q_{max}\sim 1/a_B^*$, we have
$W_{2\rightarrow 1}^{(2)}(t)\sim D^2/\rho\hbar s^3(a_B^*)^2$
from Eq. (\ref{Wif2(t)}).
More accurate calculations result in $W_{2\rightarrow 1}^{(2)}(t)=B^2(t)/2$,
see Eq. (\ref{B2(t)}). If the operation time $\tau_{op}$ is long compared to
the phonon transit time, $a_B^*/s$ ($\sim 0.3$ ps for P$_2^+$:Si), one has
from Eq. ( \ref{B2(t)}), see Ref. \cite{Fedichkin},
\begin{equation}
B^2(\tau_{op})=\frac{D^2}{3\pi^2\rho\hbar s^3(a_B^*)^2}~,
\label{B2(top)}
\end{equation}
so that the spectral function (\ref{B2(t)}) appears to be a material
constant, being about $6\cdot10^{-3}$ for the phosphorous donors in
silicon \cite{Fedichkin}, where $D=3.3$ eV, $s=9\cdot 10^5$ cm/s, and
$\rho=2.33$ g/cm$^3$. Hence,
$W_{2\rightarrow 1}^{(2)}(t)>>W_{2\rightarrow 1}^{(1)}(t)$ at
$R/a_B^*=10$ and $t<<\tilde{t}\approx 3\cdot 10^{-6}$ s, the time $\tilde{t}$
being exponentially longer for larger values of $R/a_B^*$, and in any case
longer than the operation time $\tau_{op}$, see Section III.

So, contrary to suggestions \cite{Fedichkin,Barrett} that phonon-induced
decoherence in the case of the NOT operation is determined by the value of
the relaxation rate $\Gamma_{2\rightarrow 1}$ given by Eq. (\ref{Gamma21}),
we see that at sufficiently short operation times, decoherence
in the cases of both phase and NOT operations is determined by the same
spectral function $B^2(t)$, see Eq. (\ref{B2(t)}). The distinction between
the two cases is that the diagonal elements of the density matrix remain
unchanged in the case of the phase operation since there is no relaxation,
while they decay exponentially (along with the off-diagonal matrix
elements) in the case of the NOT operation \cite{Fedichkin}.

\vskip 6mm

\centerline{\bf B. Decoherence during the auxiliary-state-assisted
operations}

\vskip 2mm

Since the excited "transport" level $|TR\rangle$ becomes temporarily
populated during the resonant-pulse operations on the P$_2^+$:Si qubit,
the phonon-induced electron transitions
$|TR\rangle\rightleftharpoons|L,R\rangle$ and
$|TR\rangle\rightleftharpoons|\chi_{1,2}\rangle$ can have a detrimental
effect on the qubit evolution, along with the transitions
$|L\rangle\rightleftharpoons |R\rangle$ and
$|\chi_1\rangle\rightleftharpoons |\chi_2\rangle$ studied above.
Let us clarify what type of the phonon-induced electron transitions
("resonant" or "non-resonant") is dominant in this
case and estimate the transition probability. We follow the line of reasoning
outlined above and start with calculations of the matrix elements
$\langle TR|e^{i{\bf qr}}|L,R\rangle$. For our choices of the "transport"
state $|TR\rangle=|\chi_3\rangle$ and the double donor orientation, see
Section II, at $R>>a_B^*$ one has $|TR\rangle\approx
\bigl[|2S\rangle_L-|2P_x\rangle_L+|2S\rangle_R+|2P_x\rangle_R\bigr]/2$,
where $\langle{\bf r}|2S\rangle_{L,R}=(8\pi (a_B^*)^3)^{-1/2}
(1-|{\bf r}-{\bf r}_{L,R}|/2a_B^*)\exp(-|{\bf r}-{\bf r}_{L,R}|/2a_B^*)$ and
$\langle{\bf r}|2P_x\rangle_{L,R}=(32\pi (a_B^*)^5)^{-1/2}
(x-x_{L,R})\exp(-|{\bf r}-{\bf r}_{L,R}|/2a_B^*)$. Neglecting the
exponentially small overlap between the localized atomic-like orbitals
centered at different donors, we have
\begin{equation}
\langle TR|e^{i{\bf qr}}|L,R\rangle=
2\sqrt{2}\frac{(qa_B^*)^2\mp i\frac{3}{2}(q_xa_B^*)}
{\left[\frac{9}{4}+(qa_B^*)^2\right]^3}e^{\mp iq_xR/2}~.
\label{TR_LR}
\end{equation}
Since, depending on the relative values of $E_L$ and $E_R$, the lowest energy
eigenstates of P$_2^+$:Si are either $|L\rangle$ and $|R\rangle$
(if $E_L\neq E_R$) or $\chi_1$ and $\chi_2$ (if $E_L=E_R$), in order to find
the probability $W_{TR}(t)$ of electron escape from the "transport" state at
$T=0$, one should add up the probabilities of, respectively,
$|TR\rangle\rightarrow |L\rangle$ and $|TR\rangle\rightarrow |R\rangle$ or
$|TR\rangle\rightarrow |\chi_1\rangle$ and
$|TR\rangle\rightarrow |\chi_2\rangle$ electron transitions. If the value of
$E_R-E_L$ is much less than the difference between $E_{TR}$ and $E_{L,R}$,
then in both cases we have the same result, so that $W_{TR}(t)$ is given by
Eq. (\ref{Wif(t)}), where now
$\omega_{if}\approx\Delta E_{31}/\hbar\approx 3E^*/8\hbar$
does not depend on $R$ at $R>>a_B^*$, and
\begin{equation}
|F_{if}({\bf q})|^2=
16\frac{(qa_B^*)^4+\frac{9}{4}(q_xa_B^*)^2}
{\left[\frac{9}{4}+(qa_B^*)^2\right]^6}|\lambda({\bf q})|^2~.
\label{Fif(q)2}
\end{equation}

Taking into account that
$q_{if}a_B^*=\omega_{if}a_B^*/s\approx 3e^2/8\varepsilon\hbar s\approx 8>>
q_{max}a_B^*\sim 1$, it is straightforward to derive from Eq. (\ref{Wif(t)})
the following expressions for the probabilities
of the "resonant" and "non-resonant" transitions, respectively,
\begin{equation}
W_{TR}^{(1)}(t)\approx \frac{8D^2}{\pi\rho\hbar s^2(a_B^*)^3}
(q_{if}a_B^*)^5 \frac{ \frac{3}{4}+(q_{if}a_B^*)^2}
{\left[\frac{9}{4}+(q_{if}a_B^*)^2\right]^6}t
\label{WTR1(t)2}
\end{equation}
and
\begin{equation}
W_{TR}^{(2)}(t)\approx\frac{176D^2}{3645\pi^2\rho\hbar s^3(a_B^*)^2
(q_{if}a_B^*)^2}~.
\label{WTR2(t)2}
\end{equation}
It follows from Eqs. (\ref{WTR1(t)2}) and (\ref{WTR2(t)2}) that
$W_{TR}^{(1)}(t)=\Gamma_{TR} t$, where
$\Gamma_{TR}\approx 3\cdot 10^7$ s$^{-1}$, and
$W_{TR}^{(2)}(t)\approx 10^{-5}$, so that the "resonant" transitions are
dominant at $t > 0.3$ ps.

Now let us check what states out of those involved in the
auxiliary-state-assisted qubit evolution are most sensitive to phonon-induced
decoherence. As we have seen above, decoherence of the low-energy states
$|L\rangle$ and $|R\rangle$ (or $|\chi_1\rangle$ and $|\chi_2\rangle$) is
quantified by the error rate  \cite{Fedichkin}, i. e., the error generated
during the operation time, $D(t)=B^2(t)/2\approx 3\cdot 10^{-3}$.
This value is greater than
$W_{TR}^{(2)}(t)$ but less than $W_{TR}^{(1)}(t)$ at $t>10^{-10}$ s, where
the processes of the spontaneous phonon emission by an electron temporarily
occupying the "transport" level become prevailing. So, at $\tau_{op}<100$ ps,
the error rate does not exceed the value of
$D(\tau_{op})\approx 3\cdot 10^{-3}$.

\vskip 6mm

\centerline{\bf V. DISCUSSION}

\vskip 2mm

Fast auxiliary-state-assisted evolution of the double-donor charge qubit
driven by the resonant electromagnetic field allows for implementation of
various one-qubit rotations in very short operation times
$\tau_{op}<100$ ps, thus minimizing the unwanted decoherence effects. At such
times, the error rate due to acoustic phonons is
$D(\tau_{op})\approx 3\cdot 10^{-3}$ at $T=0$. At finite temperatures, such
that $k_B T > \hbar\omega_0$, where $\hbar\omega_0=\hbar s/a_B^*\approx 2$
meV for dephasing processes and "non-resonant" emission/absorption
transitions, and $\hbar\omega_0=|E_i-E_f|$ for the "resonant"
$|i\rangle\rightleftharpoons |f\rangle$ transitions, the error rate increases
by a factor of $\sim k_B T / \hbar\omega_0$.

The most strong increase in the error rate at $T\neq 0$
occurs if the two donors in the
molecular ion P$_2^+$:Si are equivalent since in this case the energies $E_L$
and $E_R$ of the lowest localized states $|L\rangle$ and $|R\rangle$ are
equal to each other, and the difference $E_2-E_1$ between the eigenenergies
of the two lowest delocalized molecular states $|\chi_1\rangle$
and $|\chi_2\rangle$ is
exponentially small at large donor separations, e. g.,
$E_2-E_1\approx 10^{-6}$ meV at $R=60$ nm, see Eq. (\ref{DeltaE_21}).
To weaken decoherence,
it would be reasonable to make use of the surface gates in order to increase
the difference $E_R-E_L$ up to $E_R-E_L\sim 1$ meV so that the energy basis
of the electron be formed by the states $|L\rangle$ and $|R\rangle$ instead
of the states $|\chi_1\rangle$ and $|\chi_2\rangle$.
In this case, the electromagnetic
field should have two components driving the electron transitions
$|L\rangle\rightleftharpoons |TR\rangle$ and
$|R\rangle\rightleftharpoons |TR\rangle$ between the states $|L,R\rangle$ and
the auxiliary "transport" state $|TR\rangle$.

At $T\neq 0$, the processes of the phonon absorption by an electron
temporarily occupying the "transport" state also contribute to decoherence.
For our choice of the "transport" state, $|TR\rangle=|\chi_3\rangle$,
the state nearest to it in energy is the state $|\chi_4\rangle$.
In the case that the two
donors are equivalent and $R/a_B^*>15$, this is the $4f\sigma_u$ state
$|\chi_4\rangle\approx
\bigl[|2S\rangle_L-|2P_x\rangle_L-|2S\rangle_R-|2P_x\rangle_R\bigr]/2$
whose wave function $\langle{\bf r}|\chi_4\rangle$ is antisymmetric about the
midpoint of the line joining the two donors \cite{Fudzinaga}. At
$x=R/a_B^*>>1$ the energy separation \cite{Scott}
$E_4-E_3=E^*(x^3/4)\exp(-x/2-2)\bigl[1+O(1/x)\bigr]$ is small but greatly
exceeds the value of $E_2-E_1$, e. g., $E_4-E_3\approx 0.3$ meV at $R=60$ nm.
The donor asymmetry in the presence of the gate potentials will result in
a further increase in $E_4-E_3$, so that the phonon absorption processes
will not contribute much to decoherence at sufficiently low temperatures
$T<10$ K.

Thus, the error rate due to phonon-induced decoherence is
$D(\tau_{op})\approx 3\cdot 10^{-3}$ at $\tau_{op}<100$ ps and $T<10$ K.
This value should be compared to the error rates due to other sources of
decoherence. The lowest bounds for the decoherence times associated with the
Johnson noise from the gates and the environmental charge fluctuations are
\cite{Hollenberg,Dzurak,Barrett}, respectively, $\tau\sim 1$ $\mu$s and
$\tau\sim 1$ ns, so that the corresponding error rates \cite{Fedichkin}
$D(\tau_{op})=1-\exp(-\tau_{op}/\tau)$ do not exceed that due to phonons at
$\tau_{op}<(1\div 10)$ ps. Hence, the performance of the buried donor charge
qubit appears to be limited primarily by the electron-phonon interaction.
In this paper, we concentrated on the phosphorous donors in silicon. Since
the spectral function (\ref{B2(t)}) that ultimately determines the error
rate for one-qubit operations is a material constant, it would be worthwhile
to search for other materials and/or doping elements for the buried donor
charge qubit, in order to weaken the decoherence effects.

Although we restricted ourselves to rectangular shapes of the resonant
pulses, our consideration can be generalized to other pulse shapes
\cite{Paspalakis}. The results obtained can be also applied to quantum-dot
structures and Josephson three-level gates \cite{Paspalakis,Kis,Yang,Amin}.
Finally, once a fundamental possibility of the auxiliary-state-assisted
operations has been demonstrated, it is straightforward to organize the
coupling of P$_2^+$:Si qubits for conditional quantum operations
\cite{Hollenberg,Dzurak}.

In summary, we have proposed a scheme for fast rotations of the buried donor
charge qubit through an auxiliary-state-assisted electron evolution under
the influence of resonant microwave pulses. This scheme allows for
implementation of one-qubit operations in times as short as
$\tau_{op}\sim 1$ ps. By the example of the P$_2^+$:Si qubit, we have shown
that dephasing and "non-resonant" relaxation due to acoustic phonons are the
main sources of decoherence. The error rate at $T<10$ K and operation times
$\tau_{op}=(1\div 10)$ ps is about $3\cdot 10^{-3}$, i. e., greater than
the fault-tolerance threshold for quantum computation but sufficiently low
to investigate the small-scale devices and thus to demonstrate the
experimental feasibility of the scheme.

\vskip 2mm

\vskip 6mm

\centerline{\bf ACKNOWLEDGMENTS}

\vskip 2mm

Discussions with A. V. Tsukanov, L. Fedichkin, and M. S. Litsarev are
gratefully acknowledged.

\newpage

\includegraphics[width=\hsize]{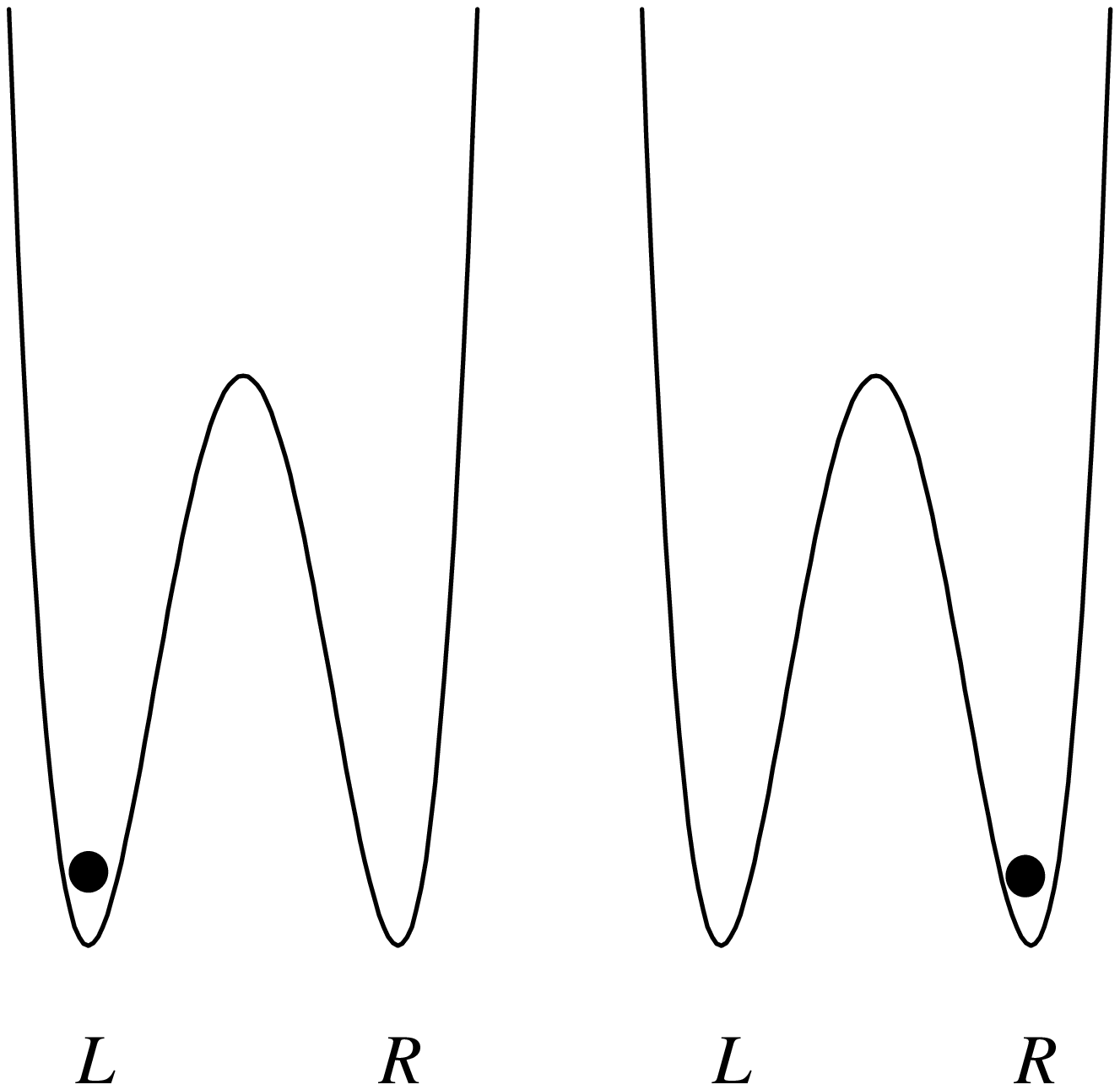}

\vskip 6mm

Fig. 1. The logical states $|0\rangle=|L\rangle$ and $|1\rangle=|R\rangle$
of the buried donor charge qubit.

\newpage

\includegraphics[width=\hsize,height=12cm]{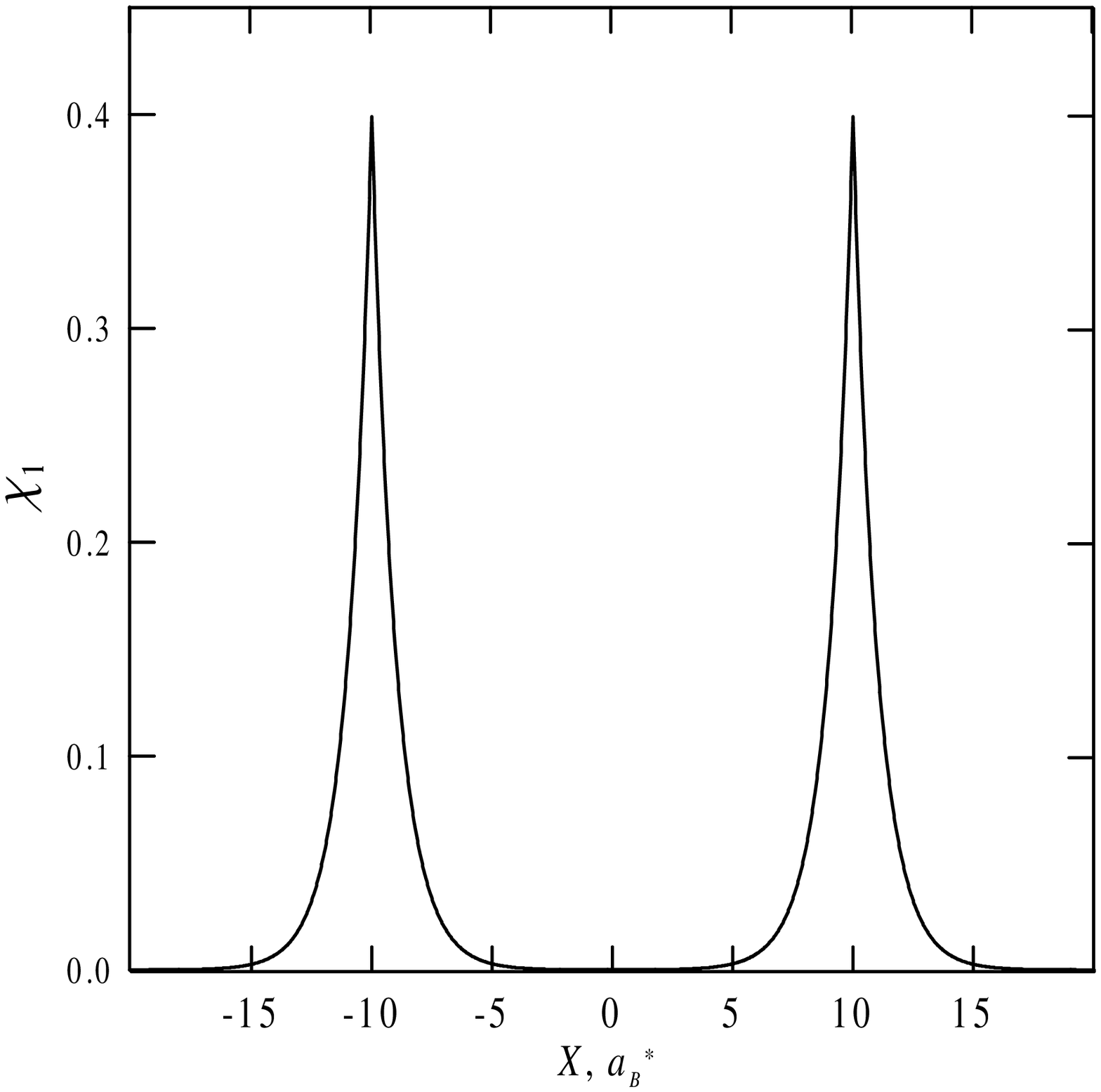}

\centerline {Fig. 2a.}

Fig. 2. One-electron wave functions of the two lowest states,
$1s\sigma_g$ (a) and $2p\sigma_u$ (b), and the excited state
$3d\sigma_g$ (c) of the molecular ion P$_2^+$:Si in the isotropic
effective mass approximation. The coordinate $X$ is along the line
joining the two donors. The donor separation is $R=20a_B^*$. The
symmetric and antisymmetric linear superpositions of $1s\sigma_g$
and $2p\sigma_u$ states correspond to $1s$ atomic states $|L\rangle$ and
$|R\rangle$ localized at the left and the right donor, respectively.
They form the qubit logical states
$|0\rangle=|L\rangle$ and $|1\rangle=|R\rangle$. The excited state
$3d\sigma_g$ is an auxiliary ("transport") state needed to transfer
an electron between $|L\rangle$ and $|R\rangle$ states upon the
influence of the external electromagnetic field.

\newpage

\includegraphics[width=\hsize]{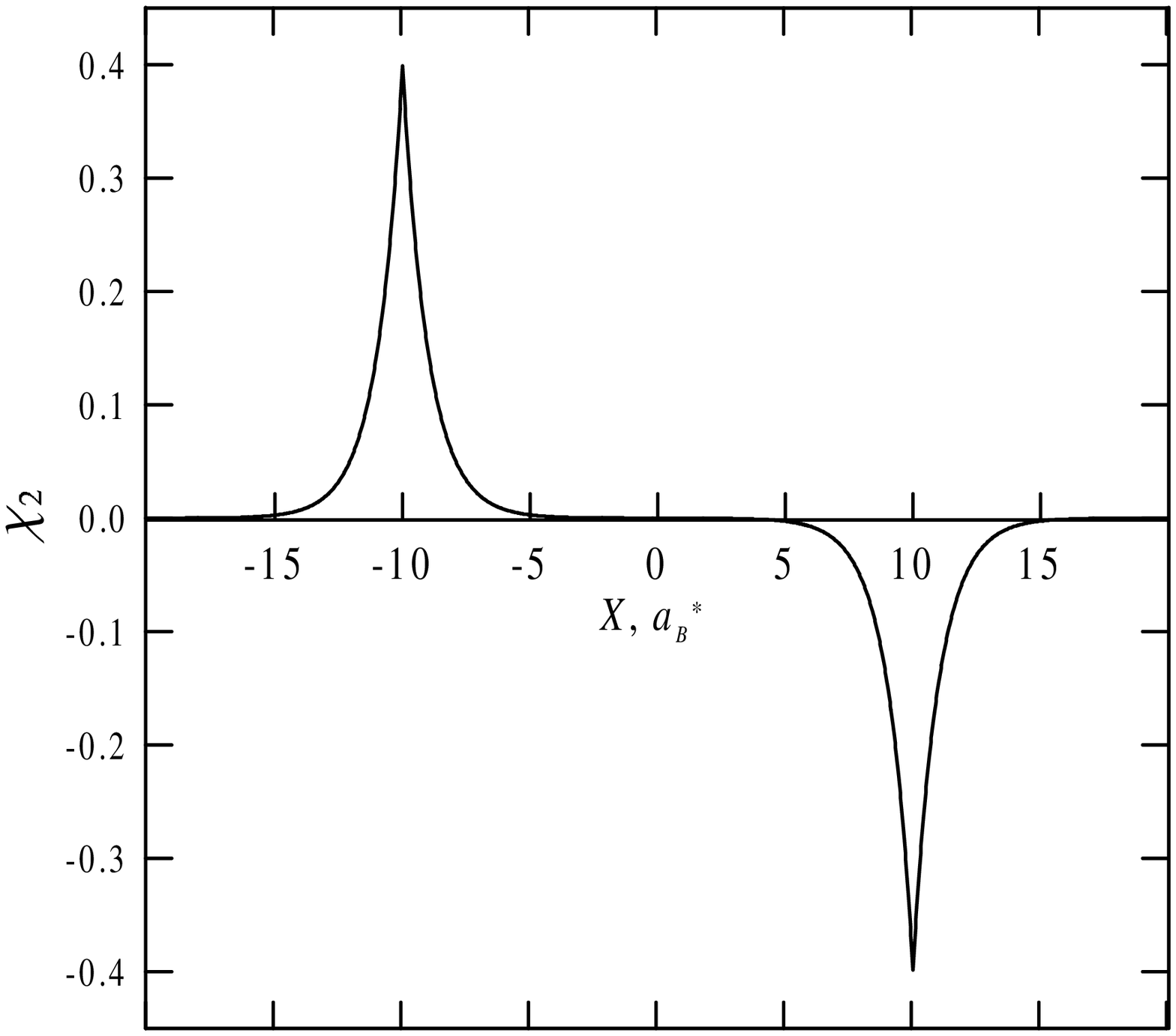}

\vskip 6mm

\centerline {Fig. 2b.}

\newpage

\includegraphics[width=\hsize]{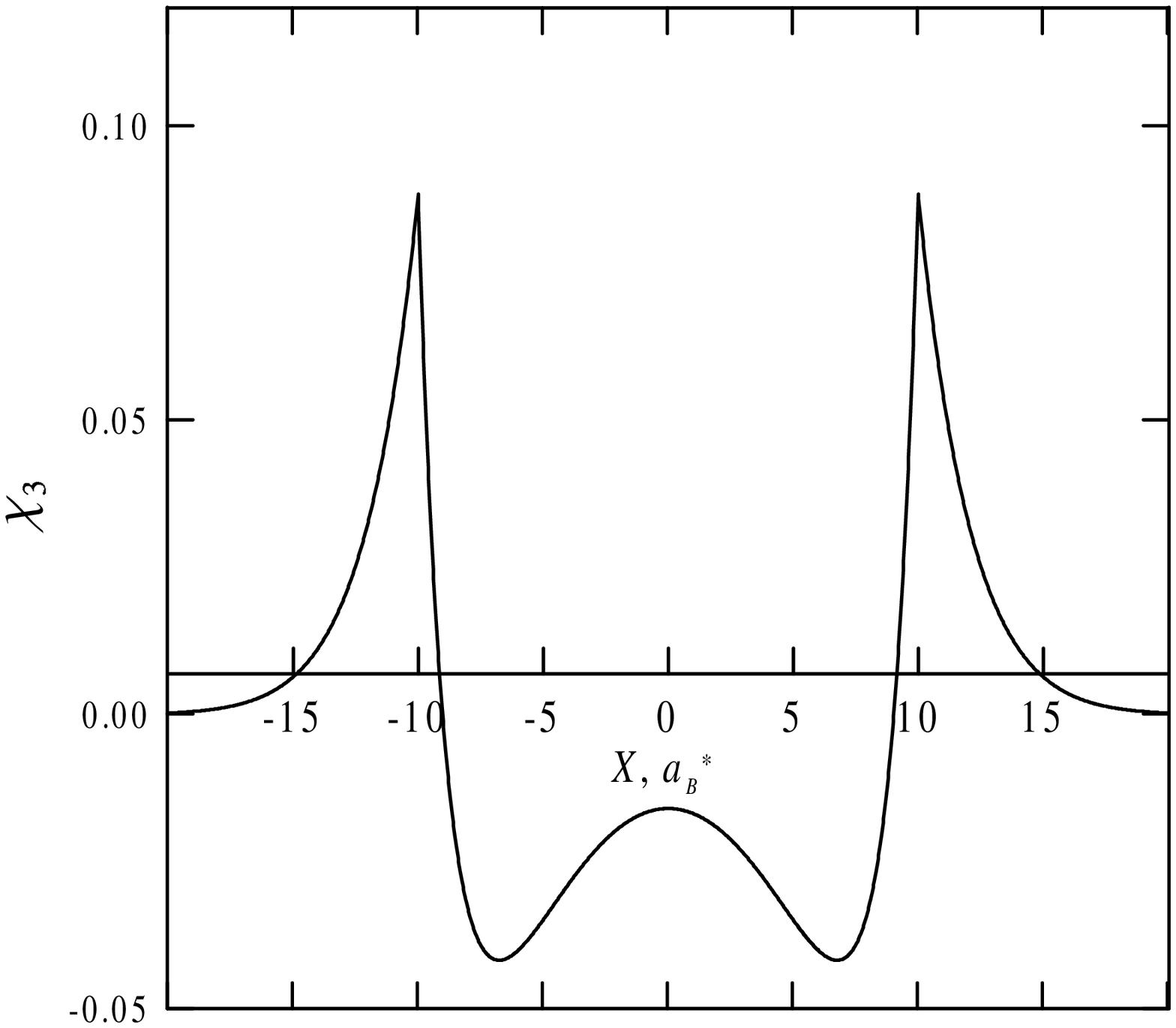}

\vskip 6mm

\centerline {Fig. 2c.}


\vskip 6mm

\begin{thebibliography}{5}

\bibitem{Barenco} A. Barenco, D. Deutsch, A. Ekert, and R. Jozsa,
Phys. Rev. Lett. {\bf 74}, 4083 (1995).

\bibitem{Shnirman} A. Shnirman, G. Schon, and Z. Hermon,
Phys. Rev. Lett. {\bf 79}, 2371 (1997).

\bibitem{Loss} D. Loss and D. P. DiVincenzo, Phys. Rev. A {\bf 57}, 120
(1998).

\bibitem{Nakamura} Y. Nakamura, Yu. A. Pashkin, and J. S. Tsai,
Nature {\bf 398}, 786 (1999).

\bibitem{Pashkin} Yu. A. Pashkin, T. Yamamoto, O. Astafiev, Y. Nakamura,
D. V. Averin, and  J. S. Tsai, Nature {\bf 421}, 823 (2003).

\bibitem{Kane} B. E. Kane, Nature {\bf 393}, 133 (1998).

\bibitem{Kane2} B. E. Kane, N. S. McAlpine, A. S. Dzurak, R. G. Clark,
G. J. Milburn, He Bi Sun, and H. Wiseman, Phys. Rev. B {\bf 61}, 2961 (2000).

\bibitem{Vrijen} R. Vrijen, E. Yablonovitch, K. Wang, H. W. Jiang,
A. Balandin, V. Roychowdhury, T. Mor, and D. DiVincenzo,
Phys. Rev. A {\bf 62}, 012306 (2000).

\bibitem{Openov} L. A. Openov, Phys. Rev. B {\bf 60}, 8798 (1999);
cond-mat/9906390.

\bibitem{Valiev} L. Fedichkin, M. Yanchenko, and K. A. Valiev,
Nanotechnology {\bf 11}, 387 (2000).

\bibitem{Tanamoto} T. Tanamoto, Phys. Rev. A {\bf 61}, 022305 (2000).

\bibitem{Oh} J. H. Oh, D. Ahn, and S. W. Hwang,
Phys. Rev. A {\bf 62}, 052306 (2000).

\bibitem{Hayashi} T. Hayashi, T. Fujisawa, H. D. Cheong, Y. H. Jeong,
and Y. Hirayama, Phys. Rev. Lett. {\bf 91}, 226804 (2003).

\bibitem{Fedichkin} L. Fedichkin and A. Fedorov,
Phys. Rev. A {\bf 69}, 032311 (2004).

\bibitem{Petta} J. R. Petta, A. C. Johnson, C. M. Marcus, M. P. Hanson,
and A. C. Gossard, e-print cond-mat/0408139.

\bibitem{Hollenberg} L. C. L. Hollenberg, A. S. Dzurak, C. Wellard,
A. R. Hamilton, D. J. Reilly, G. J. Milburn, and R. G. Clark,
Phys. Rev. B {\bf 69}, 113301 (2004).

\bibitem{Tsukanov} A. V. Tsukanov and L. A. Openov,
Fiz. Tekh. Poluprovodn. (St. Petersburg) {\bf 38}, 94 (2004)
[Semiconductors {\bf 38}, 91 (2004)].

\bibitem{Eigler} D. M. Eigler and E. K. Schweizer,
Nature {\bf 344}, 524 (1990).

\bibitem{Schofield} S. R. Schofield, N. J. Curson, M. Y. Simmons,
F. J. Rue$\beta$, T. Hallam, L. Oberbeck, and R. G. Clark,
Phys. Rev. Lett. {\bf 91}, 136104 (2003).

\bibitem{Dzurak} A. S. Dzurak, L. C. L. Hollenberg, D. N. Jamieson,
F. E. Stanley, C. Yang, T. M. Buhler, V. Chan, D. J. Reilly, C. Wellard,
A. R. Hamilton, C. I. Pakes, A. G. Ferguson, E. Gauja, S. Prawer,
G. J. Milburn, and R. G. Clark, e-print cond-mat/0306265.

\bibitem{DiVincenzo} D. P. DiVincenzo, Fortschr. Phys. {\bf 48}, 771 (2000).

\bibitem{Openov2} L. A. Openov and A. V. Tsukanov,
Pis'ma Zh. \'Eksp. Teor. Fiz. {\bf 80}, 572 (2004) [JETP Letters {\bf 80},
503 (2004)]; cond-mat/0411010;
L. A. Openov, Phys. Rev. B {\bf 70}, 233313 (2004); cond-mat/0411605.

\bibitem{Kettle} L. M. Kettle, H.-S. Goan, S. C. Smith, C. J. Wellard,
L. C. L. Hollenberg, and C. I. Pakes, Phys. Rev. B {\bf 68}, 075317 (2003).

\bibitem{Aggarwal} R. L. Aggarwal, Solid State Commun. {\bf 2}, 163 (1964).

\bibitem{Baldereschi} A. Baldereschi, Phys. Rev. B {\bf 1}, 4673 (1970).

\bibitem{Barrett} S. D. Barrett and G. J. Milburn,
Phys. Rev. B {\bf 68}, 155307 (2003).

\bibitem{Note} Note that the isotropic effective mass approximation gives
for the ground state energy of a single phosphorous donor in silicon the
value $E=-E^*/2\approx -20$ meV which is about half of the experimentally
observed value $E=-45.5$ meV, see, e. g., S. T. Pantelides and C. T. Sah,
Phys. Rev. B {\bf 10}, 621 (1974).

\bibitem{Bates} D. R. Bates and R. H. G. Reid,
Adv. At. Mol. Phys. {\bf 4}, 13 (1968).

\bibitem{Bardsley} N. Bardsley, T. Holstein, B. R. Junker, and S. Sinha,
Phys. Rev. A {\bf 11}, 1911 (1975).

\bibitem{Scott} T. C. Scott, A. Dalgarno, and J.D. Morgan, III,
Phys. Rev. Lett. {\bf 67}, 1419 (1991).

\bibitem{Fudzinaga} S. Fudzinaga, {\it Molecular orbitals method}
(Mir, Moscow, 1983), Translated from Japanese.

\bibitem{NOTE2}The resonant approximation is valid if the absolute value of
the detuning from resonance, $\hbar\delta=\hbar\omega-(E_{TR}-E_{L,R})$, is
small compared to the spacing between the energy $E_{TR}$ of the
state $|\chi_{TR}\rangle$ and the energy $E^{\prime}$ of the state
$|\chi^{\prime}\rangle$ nearest to $|\chi_{TR}\rangle$. In the case
that there are two components in ${\bf E}(t)$, the absolute values
of both $\hbar\delta_L=\hbar\omega_L-(E_{TR}-E_L)$ and
$\hbar\delta_R=\hbar\omega_R-(E_{TR}-E_R)$ should be small compared to $|E^{\prime}-E_{TR}|$.

\bibitem{Landau} L. D. Landau and E. M. Lifshitz, {\it Quantum Mechanics},
3rd ed. (Nauka, Moscow, 1974). Chapter "Perturbation theory". Note that
Eq. (\ref{aif}) is applicable for as long as
$|a_{i\rightarrow f}(\omega,t)|<<1$.

\bibitem{Bockelmann} U. Bockelmann and G. Bastard,
Phys. Rev. B {\bf 42}, 8947 (1990).

\bibitem{NOTE3} Note that at $T=0$, there are no phonons in the sample, and
hence only the second term in Eq. (\ref{Vharm}) is relevant for the
electron-phonon interaction since the initial and final phonon states,
$|i_{ph}\rangle$ and $|f_{ph}\rangle$, are, respectively, the states
$|0_{\bf q}\rangle e^{-iE_0 t}$ and
$|1_{\bf q}\rangle e^{-i(E_0+\omega_{\bf q})t}$, where
$E_0=\sum_{\bf q}(\hbar\omega_{\bf q}/2)$, so that
$\langle f_{ph}|\hat{b}^+_{\bf q}|i_{ph}\rangle=e^{i\omega_{\bf q}t}$.

\bibitem{Brandes} T. Brandes and T. Vorrath,
Phys. Rev. B {\bf 66}, 075341 (2002).

\bibitem{Kampen} N. G. van Kampen, J. Stat. Phys. {\bf 78}, 299 (1995).

\bibitem{Mozyrsky} D. Mozyrsky and V. Privman,
J. Stat. Phys. {\bf 91}, 787 (1998).

\bibitem{Liu} F.-S. Liu, K.-D. Peng, and W.-F. Chen,
Int. J. Theor. Phys. {\bf 40}, 2037 (2001).

\bibitem{Openov3} L. A. Openov, Phys. Rev. Lett. {\bf 93}, 158901 (2004);
cond-mat/0410106.

\bibitem{Paspalakis} E. Paspalakis, Z. Kis,  E. Voutsinas, and A. F. Terzis,
Phys. Rev. B {\bf 69}, 155316 (2004).

\bibitem{Kis} Z. Kis and E. Paspalakis, Phys. Rev. B {\bf 69}, 024510 (2004).

\bibitem{Yang} C.-P. Yang, S.-I. Chu, and S. Han,
Phys. Rev. A {\bf 67}, 042311 (2003).

\bibitem{Amin} M. H. S. Amin,  A. Yu. Smirnov, and A. M. van den Brink,
Phys. Rev. B {\bf 67}, 100508 (2003).

\end{thebibliography}
\end{document}